\documentclass[aps,prb,twocolumn,english,superscriptaddress,longbibliography,nobibnotes]{revtex4-2}
\usepackage{graphicx}
\usepackage[usenames,dvipsnames]{color}
\usepackage{bbm}
\usepackage{bm}
\usepackage{amsmath}
\usepackage{amssymb}
\usepackage{times}
\usepackage[colorlinks=true,linkcolor=Blue,citecolor=Blue,urlcolor=Blue]{hyperref}
\usepackage[version=3]{mhchem}

\graphicspath{{figures/compressed/}}

\newcommand{\cri}{\ce{CrI3}}

\begin{document}

\title{Exchange interactions and spin dynamics in the layered honeycomb ferromagnet \cri{}}

\author{Subhadeep Bandyopadhyay}
\thanks{These authors contributed equally to this work}
\affiliation{School of Physical Sciences, Indian Association for the Cultivation of Science, Kolkata 700 032, India}

\author{Finn Lasse Buessen}
\thanks{These authors contributed equally to this work}
\affiliation{Department of Physics, University of Toronto, Toronto, Ontario M5S 1A7, Canada}

\author{Ritwik Das}
\affiliation{School of Physical Sciences, Indian Association for the Cultivation of Science, Kolkata 700 032, India}

\author{Franz G. Utermohlen}
\affiliation{Department of Physics, The Ohio State University, Columbus, OH-43210, USA}

\author{Nandini Trivedi}
\affiliation{Department of Physics, The Ohio State University, Columbus, OH-43210, USA}

\author{Arun Paramekanti}
\affiliation{Department of Physics, University of Toronto, Toronto, Ontario M5S 1A7, Canada}

\author{Indra Dasgupta}
\affiliation{School of Physical Sciences, Indian Association for the Cultivation of Science, Kolkata 700 032, India}

\date{\today}

\begin{abstract}
We derive the microscopic spin Hamiltonian for rhombohedral \cri{} using extensive first-principles density functional theory (DFT) calculations which incorporate spin-orbit coupling and Hubbard $U$.
Our calculations indicate a dominant nearest-neighbor ferromagnetic Heisenberg exchange with weaker further-neighbor Heisenberg terms.
In addition, we find a Dzyaloshinskii-Moriya interaction which primarily drives a topological gap in the spin-wave spectrum at the Dirac point, and uncover a non-negligible antiferromagnetic Kitaev coupling between the $S=3/2$ Cr moments.
The out-of-plane magnetic moment is stabilized by weak symmetric bond-dependent terms and a local single-ion anisotropy.
Using linear spin wave theory, we find that our exchange parameters are in reasonably good agreement with inelastic neutron scattering (INS) experiments.
Employing classical Monte Carlo simulations, we study the magnetic phase transition temperature $T_c$ and its evolution with an applied in-plane magnetic field.
We further demonstrate how future high-resolution INS experiments on the magnon dispersion of single crystals in an in-plane magnetic field may be used to quantitatively extract the strength of the antiferromagnetic Kitaev exchange coupling.
\end{abstract}

\maketitle


\section{Introduction}
\label{sec:introduction}

Two-dimensional (2D) magnetic systems are attracting considerable research interest in the condensed matter community due to their ability to display unusual magnetic, electronic, and topological properties.
With the potential to realize strong coupling between magnetism and electronic or optical properties, 2D magnetic systems are also well suited to explore magneto-optical, magneto-transport, magneto-electric, or topological applications~\cite{app1,app2,app3,app4,Padmanabhan2020}. 
Yet, the formation of magnetic {\it long-range} order in 2D systems is often inhibited by the presence of thermal fluctuations, according to the Mermin-Wagner theorem~\cite{mermin}. 
In the attempt to evade the Mermin-Wagner theorem and stabilize magnetic long-range order in 2D systems, several avenues have been explored. 
Such ideas include defect engineering via vacancies or adatoms in 2D MoS$_2$ or graphene~\cite{gra1,gra2,gra3}, doping with magnetic atoms~\cite{magdope}, or placing the 2D system in proximity of a ferromagnet~\cite{app4}. 
However, a more intrinsic effect -- the presence of magnetic anisotropies which are induced by spin-orbit coupling -- can also help overcome the effect of thermal fluctuations and lead to magnetic long-range order in 2D systems. 
In this context, the discovery of magnetically ordered configurations in the few-layer limits of cleavable van der Waals (vdW) systems, such as CrX$_3$ (X=Cl,Br,I)~\cite{CrCl3,Crbr3,McGuire2015}, Cr$_2$Ge$_2$Te$_6$~\cite{CGT}, and FePS$_3$~\cite{FePS3} has stirred particular excitement. 
For \cri{}, ferromagnetic ordering can persist even in a monolayer with ordering temperature of 45~K~\cite{Huang2017}.

Bulk \cri{} has been reported to crystallize in a layered vdW structure and shows ferromagnetic ordering with T$_c$ = 61~K~\cite{Chen2018}. 
Further, a strong out-of-plane anisotropy has been observed with a band gap of 1.2 eV~\cite{McGuire2015}. 
The increased $T_c$ for \cri{} when compared to CrCl$_3$ or CrBr$_3$ further highlights the important role of the halogen ligands in magnetism. 
Heavy halogens, like iodine, can provide a strong spin-orbit coupling~(SOC) and magnetocrystalline anisotropy in the system. 
In combination with strong SOC from the ligand iodine atoms, the graphene-like honeycomb network that is formed by the magnetic Cr atoms also holds the potential to give rise to topological electronic and magnetic properties. 
In fact, recent inelastic neutron scattering~(INS) experiments on \cri{} revealed a energy gap of 2.8~meV in the spin wave spectrum at the Dirac point (`$K$'-point), suggesting the possibility of nontrivial band topology~\cite{Chen2018,Chen2021}. 
Several theoretical models have been proposed to fit the experimentally observed spin gap. 
One possible scenario is that Dzyaloshinskii–Moriya~(DM) interactions are responsible for opening up the gap, leading to a Haldane-like model for spin excitations that supports magnon Chern bands~\cite{Chen2018,Chen2020,Chen2021}. 
Alternatively, it has also been suggested that a dominant Kitaev interaction can give rise to a gap at the Dirac point and may plausibly explain experimental data~\cite{Lee2020,Chen2020,Chen2021}. 
In view of these competing proposals, the origin of the gap in the spin wave spectrum of \cri{} remains an open issue. 
It is thus important to carry out first-principles calculations to distinguish between these two scenarios for the exchange Hamiltonian for \cri{}.

To understand the mechanism that drives the formation of the spin gap, we employ density functional theory~(DFT) calculations to systematically investigate the electronic structure of \cri{} and extract effective spin exchange interactions. 
In a subsequent step, we perform Monte Carlo simulations to study the thermodynamic properties of our effective spin model and demonstrate that it can realistically capture the magnetic ordering transition, which was observed experimentally.  
In addition, we also discuss the thermodynamic properties and magnon spectrum in the presence of an in-plane magnetic field.
We show that our model (see Eqs.~\eqref{eq:model:inplane} and~\eqref{eq:model:outplane}, and Table~\ref{tab:dft:parameters}), which is derived from first-principles calculations, reproduces essential features of the magnon spectrum that are known from INS experiments. 
Finally, we demonstrate that in the presence of a large in-plane magnetic field, the momentum-dependent anisotropies of the magnon gaps at various high symmetry points directly reflect the bond-directional Kitaev interaction in \cri{} which can thus be extracted using high resolution INS experiments.

The manuscript is organized as follows. 
In Sec.~\ref{sec:crystalstructure}, we briefly outline the crystal structure of \cri{}. 
Sec.~\ref{sec:dft} is devoted to a detailed discussion of our first-principles DFT calculations for the electronic structure and the subsequent derivation of effective magnetic exchange interactions. 
We then discuss the thermodynamic properties and the spin wave spectrum of our effective spin model in Sec.~\ref{sec:model} and discuss its relevance to experimental data. 
Finally, we summarize our findings in Sec.~\ref{sec:discussion}.


\section{Crystal Structure of $\mathbf{CrI}_3$}
\label{sec:crystalstructure}

\begin{figure} 
	\centering
	\includegraphics[width=\linewidth]{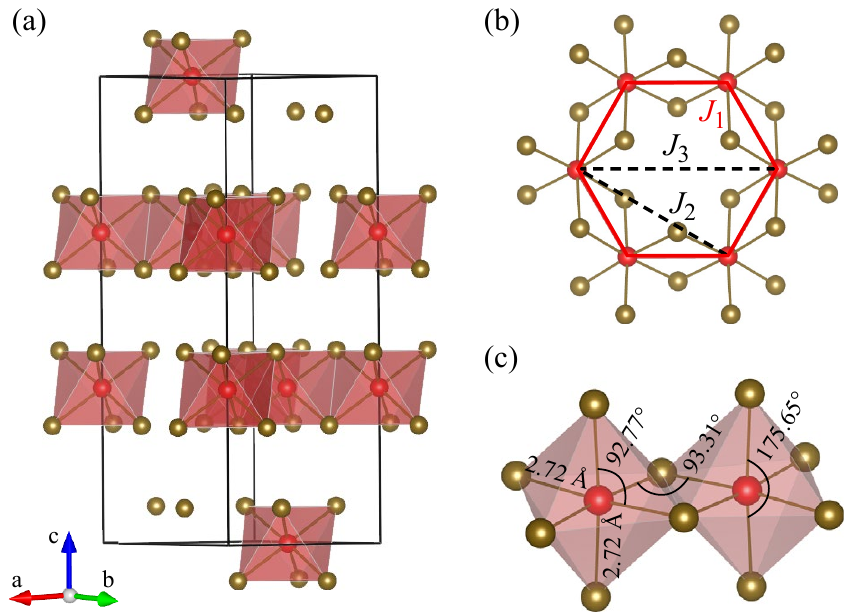}
	\caption{{\bf Crystal structure} of \cri{}. (a)~Lattice unit cell comprising of three vdW layers stacked along the $c$-direction. Cr and I atoms are colored red and brown, respectively. (b)~Honeycomb network of Cr-Cr atoms within each layer. $J_1$, $J_2$, and $J_3$ indicate Cr-Cr first, second, and third-nearest neighbor exchange paths. (c)~Geometry of edge-sharing CrI$_6$ octahedra.}
	\label{fig:crystalstructure}
\end{figure}

\cri{} crystallizes in rhombohedral $R\bar3$ (space group 148) structure with the lattice parameters $a=b=6.867~\mathrm{\AA}$ and $c=19.807~\mathrm{\AA}$~\cite{McGuire2015}. 
The edge sharing CrI$_6$ octahedral network is oriented in the $ab$-plane, forming a layered structure.
Multiple layers are stacked along the $c$-axis to form the three-dimensional crystal structure shown in Fig.~\ref{fig:crystalstructure}a. 
Within each layer, the Cr-Cr nearest neighbor bonds form a honeycomb lattice illustrated in Fig.~\ref{fig:crystalstructure}b. 
In each CrI$_6$ octahedron, the Cr-I bond lengths are equal, yet the bond angles $\angle (\text{Cr-I-Cr})$ and $\angle (\text{I-Cr-I})$ slightly deviate from an ideal octahedral structure; the average Cr-I bond length, as well as the bond angles, are displayed in Fig.~\ref{fig:crystalstructure}c. 
This slight distortion of the octahedral network is expected to affect the crystal-field splitting as we shall discuss below.


\section{Density Functional Theory Calculations}
\label{sec:dft}

To analyze the nature of magnetism in \cri{}, we perform electronic structure calculations. 
The first-principles density functional theory~(DFT) calculations are performed using the plane-wave based projector augmented wave~(PAW)~\cite{PAW} method as implemented in the Vienna ab initio simulation package (VASP)~\cite{VASP}. 
Exchange and correlation effects are treated within the generalized gradient approximation~(GGA) of Perdew-Burke-Ernzerhof~\cite{PBE}. 
To account for the effect of strong electron-electron correlation at the magnetic Cr ion, the missing correlation beyond GGA is taken into account through supplemented Hubbard $U$ (GGA+$U$) calculations~\cite{GGAU}. 
For the Hubbard $U$ we chose typical values for 3D transition metal oxides; the results reported here are obtained for $U(\mathrm{Cr})=2.7~\mathrm{eV}$ with Hunds coupling $J_H=0.7~\mathrm{eV}$. 
The kinetic energy cutoff of the plane wave basis is chosen as 500~eV and a $\Gamma$-centered $9\times9\times3$ momentum-space mesh is used for the Brillouin zone integration. 
The energy convergence criterion was set to $10^{-6}~\mathrm{eV}$ during the energy minimization process of the self-consistent cycle.

The on-site energies of the Cr-$d$ states are obtained from the muffin-tin orbital~(MTO) based $N^\mathrm{th}$ order MTO~(NMTO) method as implemented in the Stuttgart code~\cite{NMTO1,NMTO2,NMTO3}. 
We supply self-consistent potentials from the tight-binding linear muffin-tin orbital~(TB-LMTO) method for the NMTO calculations~\cite{LMTO47}. 
Space filling in the self-consistent TB-LMTO calculations within the atomic sphere approximation~(ASA) is achieved by choosing muffin-tin radii for the Cr and I atoms to be $1.46~\mathrm{\AA}$ and $1.64~\mathrm{\AA}$, respectively. 

Ultimately, we extract various exchange interactions for an effective spin model by employing the four-state method, where the required total energies for various magnetic configurations are calculated using VASP~\cite{Xiang2011,Xiang2013,Hou2018}.

\subsection{Non-spin-polarized electronic structure}

The nominal ionic formula for \cri{} is Cr$^{3+}$(I$^-$)$_3$, where the Cr$^{3+}$ ion is in the $d^3$ electronic configuration. 
Due to the formation of an octahedral network by the Cr and I ligand ions, the crystal field is expected to split the Cr-$d$ states into their $t_{2g}$ and $e_g$ manifolds. 
A small monoclinic distortion of the octahedral network, as illustrated in Fig.~\ref{fig:crystalstructure}c, further lifts the threefold degeneracy of the $t_{2g}$ states. 

\begin{figure}
	\centering
	\includegraphics[width=\columnwidth]{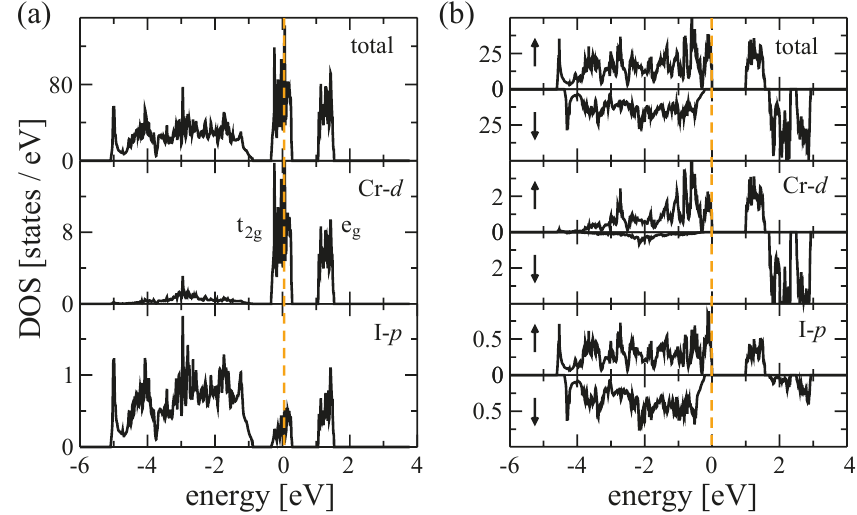}
	\caption{{\bf Density of states} in \cri{}. The Fermi energy is defined as the zero energy axis. (a)~Total DOS as well as partial DOS of the Cr-$d$ and I-$p$ states for non-spin-polarized calculations. (b)~Total and partial spin-polarized DOS. The arrows indicate the spin-up and spin-down channels.}
	\label{fig:dos}
\end{figure}

In order to verify this qualitative picture and gain more quantitative insight, we compute the non-spin polarized total density of states~(DOS) as well as the partial DOS for Cr-$d$ and ligand I-$p$ states. 
As depicted in Fig.~\ref{fig:dos}a, the partial DOS of the Cr-$d$ states shows a dominant contribution of the $t_{2g}$ manifold at the Fermi energy ($E_f$) that is hybridized with the ligands; in contrast, the $e_g$ states, which are strongly hybridized with the ligands, are completely depleted. 
The $t_{2g}$ states are half filled, which is consistent with the nominal ionic formula and lead to a metallic system. 
Our calculated value of the $t_{2g}$-$e_g$ splitting is $1.4~\mathrm{eV}$ and the bandwidth of $t_{2g}$ states is $0.65~\mathrm{eV}$.
Consequently, we retain only the Cr-$d$ orbitals in the computational basis and downfold all other orbitals using the NMTO downfolding method. 
Diagonalization of the on-site block of the corresponding real-space Hamiltonian then yields the information of crystal field splitting.
The energy eigenvalues for the Cr-$d$ states are found to be $-3.084$, $-3.061$, and $-3.061~\mathrm{eV}$ for the $t_{2g}$ states, as well as $-1.662$ and $-1.662~\mathrm{eV}$ for the $e_g$ states. 
Here, it becomes clear that the degeneracy of the $t_{2g}$ levels is indeed lifted as a consequence of the monoclinic distortion of the octahedral network. 
These distortions along with spin-orbit coupling will be crucial for the emergence of a single-ion anisotropy term and symmetric off-diagonal $\Gamma$-interactions in the model spin Hamiltonian.

\subsection{Magnetism and isotropic exchange interactions}

Our goal is to incorporate the effects of magnetism in our model; therefore, we employ spin-polarized calculations using GGA. 
Here, the ferromagnetic~(FM) configuration of Cr atoms has been considered. 
Our results for the total DOS as well as the partial DOS of the constituent atoms are displayed in Fig.~\ref{fig:dos}b. 
Within GGA, the DOS exhibits insulating behavior with fully occupied Cr-$t_{2g}$ states in the spin-up channel, whereas Cr-$t_{2g}$ states in the spin-down channel are fully depleted. 
This filling is consistent with the $d^3$ electronic configuration of the Cr$^{3+}$ ion. 
The DOS reveals a band gap of $0.98~\mathrm{eV}$ and the magnetic moment per Cr site is calculated to be $3.0~\mu_B$. 
The exchange splitting energy is found to be approximately $3~\mathrm{eV}$, which is much greater than the crystal-field splitting.

In the next step of refining our computation, we incorporate on-site Coulomb interaction and carry out GGA+$U$ calculations. 
In the FM configuration, within GGA+$U$ method, the calculated magnetic moment is $3.0~\mu_B$ and the band gap is found to be $1.2~\mathrm{eV}$. 
Both, the value of the magnetic moment and of the correlation induced gap, are in agreement with data reported in Ref.~\cite{McGuire2015}. 
To identify the electronic ground state, besides the FM configuration, we have considered additional possible magnetic configurations within the unit cell as follows: (i)~intra-layer FM in combination with inter-layer antiferromagnetic (AFM) configuration of Cr spins, and (ii)~simultaneous intra-layer and inter-layer AFM configuration of Cr spins. 
Our total energy calculations reveal that among those three configurations the FM configuration has the lowest energy.

In order to determine the effective symmetric Heisenberg magnetic exchange interaction between the Cr atoms, we employ the four state method, which allows us to extract the exchange constant based on the energies of four distinct spin configurations~\cite{Xiang2011,Xiang2013}. 
For a particular pair of Cr ions, which we refer to as $i$ and $j$, we consider the following spin configurations: (i)~spin up at site $i$ and spin up site $j$, (ii)~spin up at site $i$ and spin down at site $j$, (iii)~spin down at site $i$ and spin up at site $j$, and (iv)~spin down at site $i$ and spin down at site $j$. 
In every one of the four configurations, we keep the spin of all other Cr sites fixed. 
In the following, we assume a Heisenberg spin Hamiltonian of the form $H = \sum_{ij} J_{ij} \mathbf{S}_i \mathbf{S}_j$, where the sum runs over arbitrary pairs of Cr ions $i$ and $j$; in practice, however, we constrain the sum to only include terms up to third-nearest neighbor sites. 
Note that in our notation $J_{ij}<0$ amounts to FM coupling and $J_{ij}>0$ indicates AFM interaction. 
Using VASP, we calculate the energies $E_1$, $E_2$, $E_3$ and $E_4$, respectively, for the four spin configurations (i)--(iv). 
The exchange interaction $J_{ij}$ is then calculated as $J_{ij} = (E_1-E_2-E_3+E_4)/4S^2$, where S=3/2 for Cr$^{3+}$. 

We compute several different Heisenberg exchange constants, three of which are within the honeycomb plane. 
We consider nearest neighbor interaction $J_1$, second-nearest neighbor interaction $J_2$, and third-nearest neighbor interaction $J_3$; in addition, we compute the three distinct inter-layer exchanges $J_{c1}$, $J_{c2}$, and $J_{c3}$. 
See Fig.~\ref{fig:crystalstructure}b for the definition of in-plane couplings and Fig.~\ref{fig:model:geometry}c for the inter-plane couplings. 
Our calculations reveal strongly FM intra-layer interaction $J_1=-2.9~\mathrm{meV}$ and a weaker AFM inter-layer exchange $J_{c1}=0.1~\mathrm{meV}$, as well as FM exchanges $J_{c2}=-0.15~\mathrm{meV}$ and $J_{c3}=-0.22~\mathrm{meV}$.
The further neighbor in-plane interactions are found to be FM $J_2=-0.3~\mathrm{meV}$ and AFM $J_3=0.2~\mathrm{meV}$.
These calculated parameters are consistent with the findings in previous INS studies~\cite{Chen2020}.
The FM nature of the dominant NN exchange interaction $J_1$ can further be inferred from the superexchange mechanism between the Cr sites, as the dominant Cr-I-Cr exchange paths form an angle of approximately $90^\circ$ (see Fig.~\ref{fig:crystalstructure}c).


\subsection{Effect of spin orbit coupling}

The large atomic mass of iodine indicates that the inclusion of SOC in our first-principles calculations is important to correctly predict the magnetic properties of \cri{}.
Upon including SOC, our calculations show that the FM order in the system is further stabilized, and the total energy is further lowered. 
The total magnetic moment is calculated to be $3.0~\mu_B$ per Cr ion.
In addition, the Cr ions also gain a substantial orbital magnetic moment of $0.07~\mu_B$, indicating the strong effect of SOC. 
Unlike our previous calculations in the absence of SOC, we now find that there is an anisotropy in the FM alignment, which we quantify via the anisotropy energy $E_\mathrm{aniso}$ that is defined as the difference between the energy of the in-plane FM configuration and that of the out-of-plane configuration. 
For \cri{}, we find that $E_\mathrm{aniso}=0.3~\mathrm{meV}$ per Cr ion. 
Consequently, magnetic moments are predicted to favor an out-of-plane FM alignment, which is in agreement with reported results~\cite{Chen2020,McGuire2015}.

Capturing the anisotropy in the magnetic configuration requires a more complicated effective spin model, which goes beyond simple Heisenberg interactions. 
We therefore assume the generalized symmetry-allowed 
microscopic spin Hamiltonian of the form $H=\sum_{ij,\alpha\beta} J^{\alpha \beta}_{ij} S^\alpha_i S^\beta_j$, where the Cr ions on sites $i$ and $j$ can now couple through arbitrary components $\alpha,\beta=x,y,z$ of the $S=3/2$ spin operators.
The generalized $3\times 3$ interaction matrices $J^{\alpha \beta}_{ij}$ are often expressed in terms of Heisenberg interaction $J$, Kitaev interaction $K$, symmetric off-diagonal $\Gamma$-interaction, and Dzyaloshinskii-Moriya interaction $D$; furthermore, we consider a local magnetic anisotropy $A$. 
The explicit form of the effective spin Hamiltonian, where we retain terms up to third-nearest neighbors, is discussed in the following Sec.~\ref{sec:model}.

\begin{table}
	\caption{{\bf Exchange constants} for the microscopic model of \cri{} obtained from our {\it ab initio} calculations~(DFT) as detailed in Sec.~\ref{sec:dft}, as well as experimental values extracted from inelastic neutron scattering~(INS) data in Ref.~\cite{Chen2021}. A dash signals that the respective parameter has not been included in the modeling of the experimental data. All exchange constants are given in units of meV.}
	\centering
	\begin{tabular}{l r r}
	\hline\hline
	 & DFT & INS \\
	\hline
	$J_1$ & $-2.70$ & $-2.11$ \\
	$J_2$ & $-0.30$ & $-0.11$ \\
	$J_3$ & $0.24$ & $0.10$ \\
	$K$ & $0.60$ & -- \\
	$\Gamma$ & $-0.12$ & -- \\
	$D$ & $0.22$ & $0.17$ \\
	$A$ & $-0.13$ & $-0.12$ \\
	$J_{c1}$ & $0.09$ & $0.05$ \\
	$J_{c2}$ & $-0.15$ & $-0.07$ \\
	$J_{c3}$ & $-0.22$ & $-0.07$ \\
	\hline\hline
	\end{tabular}
	\label{tab:dft:parameters}
\end{table}

We calculate all relevant exchange interactions for our effective model using the four-state method and choosing appropriate spin configurations~\cite{Hou2018}. 
By far the most dominant energy scale in the system is set by the FM nearest neighbor Heisenberg interaction $J_1=-2.7~\mathrm{meV}$. 
All remaining exchange couplings are found to be significantly smaller; the complete set of exchange constants is listed in Table~\ref{tab:dft:parameters}.
Generally, the isotropic Heisenberg terms compare well with our previous estimate based on spin-polarized calculations in the absence of SOC.
The Kitaev interaction is found to be small but antiferromagnetic; this suggests the important role of SOC on iodine.
In addition, the $\Gamma$-interaction and local magnetic anisotropy $A$, which arise from the distortion of the CrI$_{6}$ octahedra and SOC, ensure that the FM ground state configuration is oriented perpendicular to the honeycomb planes. 
We however note that while the anisotropic interaction constants are nonzero, and important to ensure a nonzero $T_c$ in the \cri{} monolayer, their magnitudes are small and therefore subject to some numerical uncertainty.
While the stability of FM correlations is set by the dominant energy scale $J_1$ and is expected to be numerically sound, the pinning of magnetic moments to the out-of-plane direction is set by sub-leading energy scales $\Gamma$ and $A$. 
Interestingly, we find that the nearest neighbor inter-layer Heisenberg exchange interaction is antiferromagnetic while the next-nearest neighbor interaction between adjacent layers is ferromagnetic. 
This is in good agreement with experimental values extracted from inelastic neutron scattering~(INS) data in Ref.~\cite{Chen2021}. 

The bond dependent antiferromagnetic (AFM) Kitaev interaction obtained in our calculations suggests the importance of the $e_{g}$-$t_{2g}$ paths in mediating the exchange interaction as shown using a perturbative calculation on a tight-binding model~\cite{hykee2021}. 
The $e_{g}$-$t_{2g}$ path is promoted by the I ligands subject to strong SOC, where the I p-states strongly hybridize with the $e_{g}$ states of Cr.
However, the antiferromagnetic Kitaev exchange $K/J \!\sim\! 0.2$, which we obtain from our {\it ab initio} calculations, is significantly larger than the previously estimated value $K/J \!\sim\! 0.01$ using perturbation theory. 
The difference may stem from the estimate of the on-site energies  and hopping parameters employed in Ref.~\cite{hykee2021}, which can impact the balance between competing pathways with opposite signs for the Kitaev interaction. 
Furthermore, this previous work did not include interactions beyond nearest neighbor lattice sites and was thus unable to address the relevance of DM interaction on next-nearest neighbor bonds.

In the following section, we perform a detailed analysis of the effective spin model, both with and without the application of an external magnetic field.


\section{Effective spin model}
\label{sec:model}

Informed by the properties of \cri{} that we uncovered in our DFT calculations, our complete microscopic spin model on a layered honeycomb lattice includes Heisenberg, Kitaev, $\Gamma$, and Dzyaloshinskii-Moriya (DM) spin exchange, as well as a local single-ion anisotropy. 
Our model is captured by the Hamiltonian $H=H_\parallel+H_\perp$, which is parametrized by the set of exchange constants ($J_1$,$J_2$,$J_3$,$K$,$\Gamma$,$D$,$A$,$J_{c1}$,$J_{c2}$,$J_{c3}$); the first therm, $H_\parallel$, denotes in-plane interactions while the second therm, $H_\perp$, resembles inter-plane exchange. 
The two terms are given by 
\begin{align}
\label{eq:model:inplane}
H_\parallel&=
\!\!\!\!\! \sum\limits_{\langle i,j \rangle_{\gamma\neq\alpha,\beta}} \!\!\!\!\! \Big( J_1~\mathbf{S}_i \mathbf{S}_j + K~\widetilde{S}^\gamma_i \widetilde{S}^\gamma_j + \Gamma~( \widetilde{S}^\alpha_i \widetilde{S}^\beta_j + \widetilde{S}^\beta_i \widetilde{S}^\alpha_j ) \Big) \nonumber\\
&+ \!\! \sum\limits_{\langle\langle i,j \rangle\rangle} \!\!  \Big( J_2~\mathbf{S}_i \mathbf{S}_j + D~ \widehat{\mathbf{d}}_{ij} \cdot \mathbf{S}_i \times \mathbf{S}_j \Big) + \!\!\!\! \sum\limits_{\langle\langle\langle i,j \rangle\rangle\rangle} \!\!\!\! J_3~\mathbf{S}_i \mathbf{S}_j \nonumber\\
&+ \sum\limits_{i} A~S_i^z S_i^z 
\end{align}
and 
\begin{align}
\label{eq:model:outplane}
H_\perp&= 
\sum\limits_{\langle i,j \rangle_\perp} \!\! J_{c1}~\mathbf{S}_i \mathbf{S}_j 
+ \!\!\!\! \sum\limits_{\langle\langle i,j \rangle\rangle_\perp} \!\!\!\! J_{c2}~\mathbf{S}_i \mathbf{S}_j
+ \!\!\!\!\! \sum\limits_{\langle\langle\langle i,j \rangle\rangle\rangle_\perp} \!\!\!\!\!\! J_{c3}~\mathbf{S}_i \mathbf{S}_j \,.
\end{align}
\begin{figure}
	\centering
	\includegraphics[width=\linewidth]{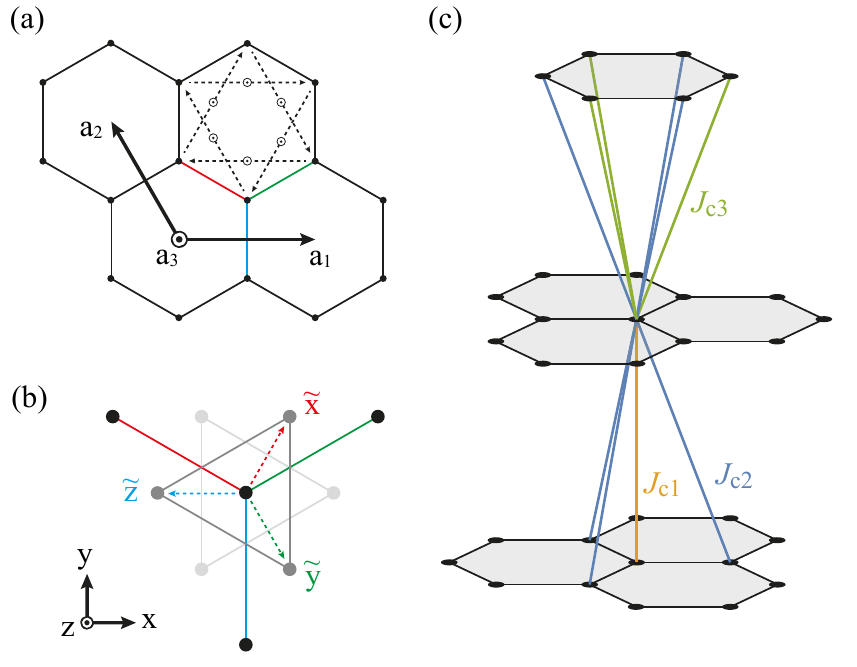}
	\caption{{\bf Layered honeycomb model} for \ce{CrI3}. (a)~Convention for in-plane interactions: red (green, blue) bonds denote $x$-type ($y$, $z$-type) bonds. Dashed arrows from site $i$ to $j$ indicate the orientation of DM interactions $\widehat{\mathbf{d}}_{ij} \cdot \mathbf{S}_i \times \mathbf{S}_j$ with $\widehat{\mathbf{d}}_{ij}$ pointing in the $\mathbf{a}_3$ direction. (b)~Cr-atoms (black) surrounded by I-atoms (top layer dark gray, bottom layer light gray). The local coordinate system for Kitaev-like interactions is spanned by $\widetilde{x}$, $\widetilde{y}$, and $\widetilde{z}$. The laboratory frame is spanned by $x$, $y$, and $z$. (c)~Layering of honeycomb sheets with inter-plane couplings. The yellow lines indicate inter-plane nearest neighbor coupling $J_{c1}$; blue and green lines denote $J_{c2}$ and $J_{c3}$, respectively. }
	\label{fig:model:geometry}
\end{figure}
The first sum runs over nearest neighbor bonds of type $\gamma=x,y,z$ within the honeycomb layers, with $\alpha,\beta \neq \gamma$ denoting the remaining two bond types. 
The sums over $\langle\langle i,j \rangle\rangle$ and $\langle\langle\langle i,j \rangle\rangle\rangle$ run over second-nearest and third-nearest neighbors within the honeycomb planes, respectively; similarly, sums indicated by $\langle i,j \rangle_\perp$, $\langle\langle i,j \rangle\rangle_\perp$, and $\langle\langle\langle i,j \rangle\rangle\rangle_\perp$ run over inter-plane nearest, second-nearest, and third-nearest neighbors; there exists one inter-plane nearest neighbor, six inter-plane second-nearest neighbors, and three inter-plane third-nearest neighbors for every Cr ion. 
Spin operators $\mathbf{S}_i=(S_i^x,S_i^y,S_i^z)$ represent S=3/2 moments with components in the laboratory frame (the $xyz$-frame indicated in Fig.~\ref{fig:model:geometry}b). 
Rotated spin operators $\widetilde{S}_i=(\widetilde{S}_i^x,\widetilde{S}_i^y,\widetilde{S}_i^z)$ in the Kitaev and $\Gamma$ interaction terms are written in the local basis, i.e. the $\widetilde{x}\widetilde{y}\widetilde{z}$-frame illustrated in Fig.~\ref{fig:model:geometry}b. 
The (unit length) DM vectors $\widehat{\mathbf{d}}_{ij}$ are aligned in the $\pm z$ direction, with their sign structure as shown in Fig.~\ref{fig:model:geometry}a. 
Note that the nearest neighbor DM interactions vanish by symmetry, while second-nearest neighbor DM interactions are allowed with an out-of-plane DM vector~\cite{Chen2018}. 
We have also computed a symmetry-allowed in-plane DM term and find it to be negligible.
In the following subsections, we study the thermal and dynamic properties of the model Hamiltonian. 
For the remainder of the manuscript, we shall focus on the two-dimensional magnetism of the monolayer, neglecting the inter-plane exchanges in our model Hamiltonian.


\subsection{Magnetic phase diagram}
\label{sec:mc}

In order to study the stability of the ferromagnetically ordered ground state of \cri{} in the presence of thermal fluctuations, we perform classical Monte Carlo simulations of our model Hamiltonian Eq.~\eqref{eq:model:inplane}. 
We simulate systems of $32\times32$ unit cells ($N=2048$ spins in total) with periodic boundary conditions in the temperature range from 5~K to 80~K. 
The simulation of 144 replicas at logarithmically spaced temperature points is performed in a parallel tempering scheme, which accelerates the convergence of the simulations~\cite{Katzgraber2006}. 

From the Monte Carlo simulations -- based on the exchange constants obtained in our DFT calculations -- we find that \cri{} orders ferromagnetically below a critical temperature of $T_c^\mathrm{2D}=40.6~\mathrm{K}$, and the magnetic moment of the ordered state lies perpendicular to the honeycomb plane. 
The value of the critical temperature is in good agreement with the transition temperature $T_{c,\mathrm{expt}}^\mathrm{2D}=45~\mathrm{K}$ that has been determined experimentally for monolayer \cri{} samples~\cite{Huang2017}. 

\begin{figure}
	\centering
	\includegraphics[width=\linewidth]{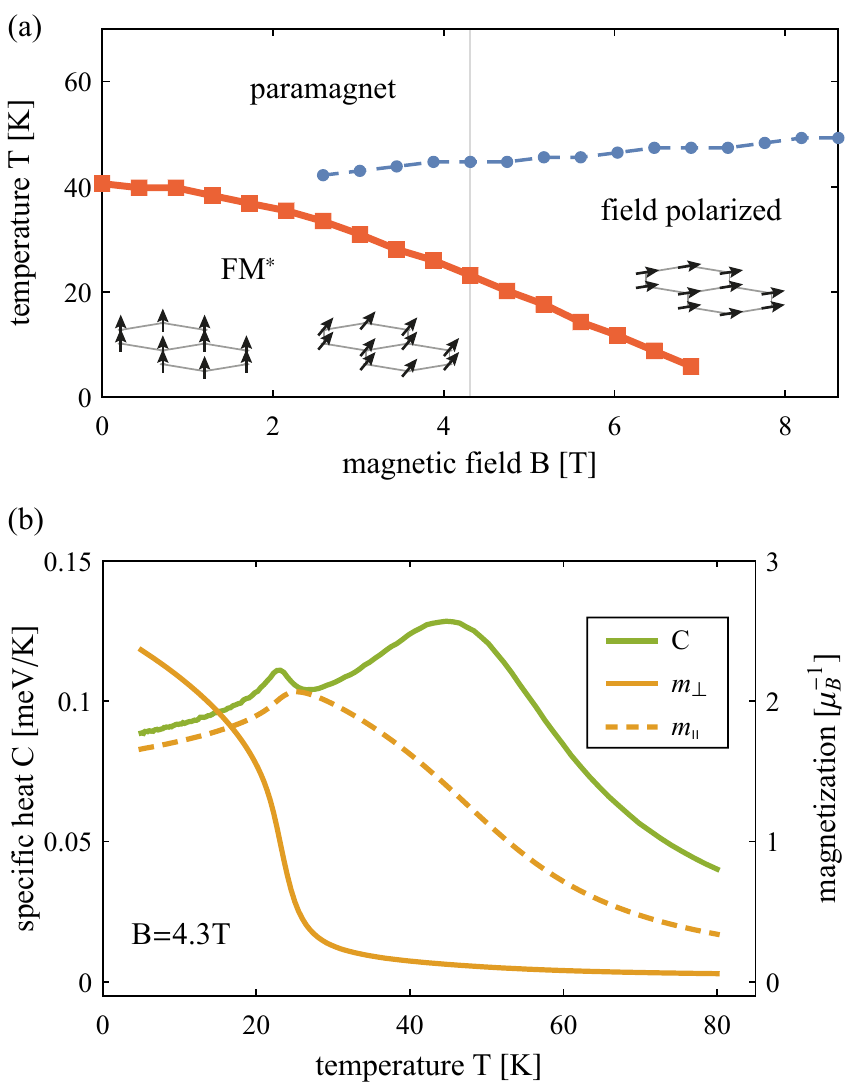}
	\caption{{\bf Phase diagram} of \ce{CrI3}. (a)~Magnetic phase as a function of temperature and external magnetic field $B$. The dashed blue line marks a crossover from the paramagnetic state to the field-polarized state. The solid line indicates the transition into the ferromagnetically ordered state~FM$^*$ with finite out-of-plane magnetization. The spontaneous polarization in the FM$^*$ phase tilts continuously from the out-of-plane direction to the in-plane direction as the magnetic field is increased. The respective spin configurations are illustrated in the insets. (b)~Specific heat and magnetization as a function of temperature at $B=4.3~\mathrm{T}$, as indicated by the gray line in subpanel~(a). Results are obtained in Monte Carlo simulations with model parameters extracted from our DFT calculations. Statistical error bars are smaller than the line width.}
	\label{fig:mc:phasediagram}
\end{figure}

Next, we investigate the thermal properties of \cri{} subject to an external magnetic field $B$, which is applied within the honeycomb planes in the $[\overline{1} 1 0]$ direction, i.e., along the armchair direction of $x$-type bonds, see the geometry illustrated in Figs.~\ref{fig:model:geometry}a and~\ref{fig:model:geometry}b. 
To this end, we add a field-coupling term $H_B = g \mu_B B \sum_i \hat{\mathbf{n}}_{\overline{1} 1 0} \mathbf{S}_i$ to the Hamiltonian Eq.~\eqref{eq:model:inplane}, where $\hat{\mathbf{n}}_{\overline{1} 1 0}$ is the unit vector pointing in the $[\overline{1} 1 0]$ direction and $g=2$. 
The resulting phase diagram as a function of temperature and magnetic field strength is summarized in Fig.~\ref{fig:mc:phasediagram}a, and it reveals the existence of three distinct phases: 
At high temperatures, the system naturally is a thermally disordered paramagnet. 
In the presence of a strong external magnetic field, a field-polarized state manifests in which the magnetic moment is  aligned with the in-plane field with no out-of-plane magnetization.
The third phase, as discussed in the previous paragraph, is observed when the system undergoes a phase transition into its low-temperature ferromagnetically ordered phase in the absence of an external field. 
In a nonzero in-plane field, this phase still exhibits canted ferromagnetic order, with the global magnetization no longer perpendicular to the honeycomb plane; instead, it gradually tilts from an out-of-plane orientation to an in-plane direction as the magnetic field strength is increased. 
We refer to this canted ferromagnetic ordered phase as FM$^*$ order. 

At zero temperature, the critical magnetic field which drives the transition from the FM$^*$ ordered state into the field-polarized phase is found to be $B_c^\mathrm{2D}\approx 7~\mathrm{T}$. 
As such, the value exceeds the experimentally determined $B_{c,\mathrm{expt}}^\mathrm{2D}=3.5~\mathrm{T}$ approximately by a factor of two~\cite{McGuire2015}. 
However, we emphasize that $B_c$ is determined only by the subleading exchange constants $\Gamma$ and $A$, which pin the magnetic moment along the out-of-plane direction. 
Since the values of these parameters are only approximately 5\% of the leading energy scale $J_1$ in our DFT calculations, they may be subject to sizable relative uncertainty; consequently, the numerical prediction for the critical magnetic field should be interpreted with caution. 

The global phase diagram displayed in Fig.~\ref{fig:mc:phasediagram}a implies that there exists a two-step ordering process for a finite window of intermediate field strength $3~\mathrm{T}\lesssim B \lesssim 7~\mathrm{T}$. 
With decreasing temperature, the system gradually builds up a finite in-plane magnetization $m_\parallel = g\mu_B \sum_i \hat{\mathbf{n}}_{\overline{1} 1 0} \mathbf{S}_i$ as the crossover into the field-polarized state is approached, see the exemplary data for specific heat and magnetization in Fig.~\ref{fig:mc:phasediagram}b for an external magnetic field strength $B=4.3~\mathrm{T}$.
The crossover is further signaled by a broad maximum in the specific heat around $T=45~\mathrm{K}$. 
It is noteworthy, however, that the crossover is not associated with the buildup of any out-of-plane magnetization $m_\perp = g\mu_B \sum_i \hat{\mathbf{n}}_{0 0 1} \mathbf{S}_i$. 
The latter only occurs at a lower temperature scale $T=23~\mathrm{K}$ and is accompanied by a sharper peak in the specific heat. 
Since the data shown in Fig.~\ref{fig:mc:phasediagram}b is at intermediate field strength, the magnetic moment in the ground state configuration is neither fully aligned in-plane nor out-of-plane, i.e., both $m_\parallel$ and $m_\perp$ assume finite values at low temperature but do not saturate. 
We further mention that the specific-heat signature of the lower temperature transition into the ground state configuration changes with the applied field strength: At low field, when the phase transition is associated with the largest possible reconfiguration of the magnetic moment (from fully in-plane to fully out-of-plane), the concomitant peak in the specific heat is most distinct. 
At increasing field strength, when the shift in the magnetic moment becomes smaller (i.e., the ground state magnetization is no longer fully out-of-plane), the signature in the specific heat is observed to gradually become less pronounced until it disappears entirely above $B_c^\mathrm{2D} \approx 7~\mathrm{T}$.


\subsection{Magnon band structure}
\label{sec:sw}

Having explored the ground state properties of \cri{} in the previous section, we now turn to the excitation spectrum of the system. 
We perform linear spin wave calculations in order to unveil magnon excitations, which may exist on top of the ferromagnetically ordered ground state. 
The spin wave calculations are performed on the two-site magnetic unit cell depicted in Fig.~\ref{fig:model:geometry}a, resulting in two distinct magnon bands. 

\begin{figure}
	\centering
	\includegraphics[width=\linewidth]{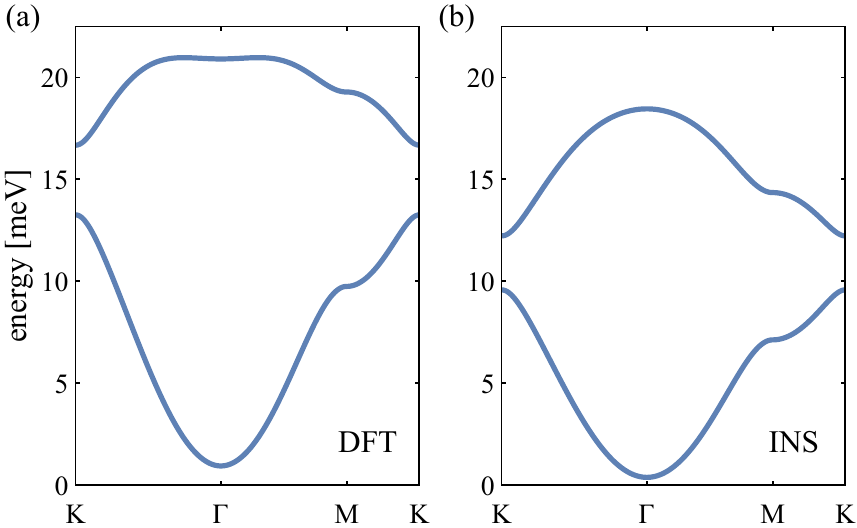}
	\caption{{\bf Spin wave spectrum} of the microscopic model for \ce{CrI3} with exchange constants (a)~obtained from {\it ab initio} calculations and (b)~fitted to experimental inelastic neutron scattering data.}
	\label{fig:sw:spinwave_composite}
\end{figure}

In a first step, we compute the spin wave spectrum for \cri{} with the exchange constants obtained from our DFT calculations. 
Our theoretical prediction for the spectrum reproduces key features that have previously been unveiled in experiments. 
In particular, a band gap of approximately 3.5~meV at the Brillouin zone (BZ) corner is observed in addition to a 0.9~meV gap at the BZ center and an overall bandwidth of approximately 20~meV. 
These predictions compare well to the experimental numbers for the spin gap at the BZ center, 0.3--1~meV~\cite{Chen2018,Chen2021,Lee2020}, and at the BZ corners, 4~meV~\cite{Chen2018}. 
However, deviations from the experimental data are also observed. 
The detailed spectrum, which we predict based on our DFT calculations for the exchange constants, is shown in Fig.~\ref{fig:sw:spinwave_composite}a, plotted along a high-symmetry path from the Brillouin zone corner ($K$-point) via the BZ center ($\Gamma$-point), the middle of the BZ edge ($M$-point), and back to the BZ corner. 
For reference, the spin wave spectrum that is obtained from exchange constants fitted to best reproduce experimental inelastic neutron scattering (INS) data in Ref.~\cite{Chen2021} is displayed in Fig.~\ref{fig:sw:spinwave_composite}b. 
Note that the data shown here is for in-plane interactions only, i.e., we neglect the inter-plane interactions extracted from the experiment. 
The direct comparison shows that the Dirac gap at the $K$-points lies at slightly increased energy levels and the upper band is significantly flattened around the BZ center. 
We further address these deviations in the next subsection.


\subsection{Comparison with experimental data}

\begin{figure}
	\centering
	\includegraphics[width=\linewidth]{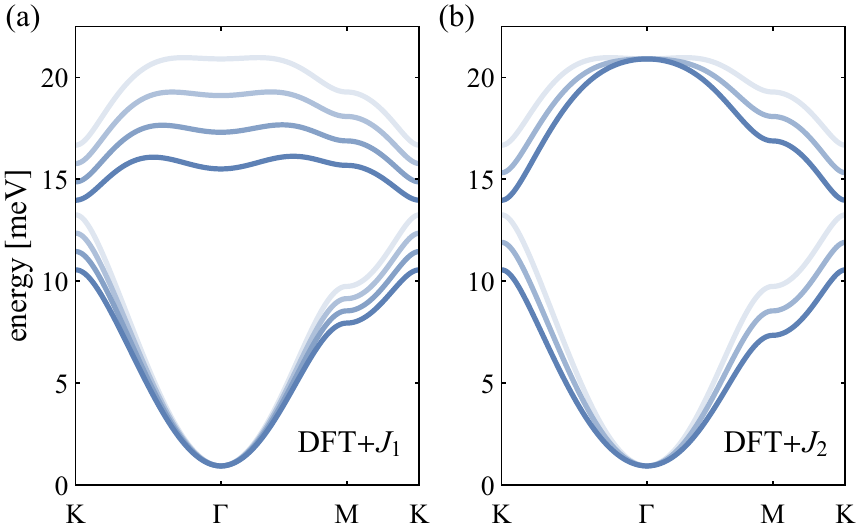}
	\caption{{\bf Modified spin wave spectrum} under deformed Heisenberg interactions $J_1$ and $J_2$, based on our DFT calculations. (a)~Nearest neighbor interactions $J_1=-2.7$, $-2.5$, $-2.3$, $-2.1~\mathrm{meV}$ (from light to opaque color) control the overall band width. (b)~Next-nearest neighbor interactions $J_2=-0.3$, $-0.2$, $-0.1~\mathrm{meV}$ (from light to opaque color) generally lower the bands, except at the Brillouin zone center.}
	\label{fig:sw:spinwave_DFT_modified_composite}
\end{figure}

We now turn to a more detailed discussion of the differences between the spin wave spectrum predicted from our DFT calculations and the spin wave spectrum extracted from inelastic neutron scattering experiments. 
We identify two salient differences: (i)~The predicted Dirac gap is shifted towards higher energy levels and (ii)~the upper band is significantly flattened around the BZ center. 
The most direct way to tune the overall band width -- and hence the energy level at which the Dirac gap is observed -- is to alter the leading energy scale in the model, which in our case is the nearest neighbor Heisenberg interaction $J_1$. 
The effect of varying $J_1$ is illustrated in Fig.~\ref{fig:sw:spinwave_DFT_modified_composite}a, where we plot the band structure based on our DFT calculations, but with modified exchange constant $J_1$. 
While a reduction of $J_1$ indeed reduces the overall bandwidth and shifts the Dirac gap down to lower energies, solely tuning the leading energy scale does not remedy the flatness of the upper band. 
Rather, the flatness of the upper band is tied to the next-nearest neighbor exchange coupling $J_2$; small changes of the latter can have a strong impact on the shape of the spin wave spectrum, as illustrated in Fig.~\ref{fig:sw:spinwave_DFT_modified_composite}b. 
In fact, a reduction from $J_2=-0.3~\mathrm{meV}$ to $J_2=-0.1~\mathrm{meV}$ already brings the predicted spin wave spectrum much closer to the experimental findings. 
Such a change of less than 10\% of the principal energy scale in our model can be expected to be within the uncertainty of DFT calculations. 

\begin{figure}
	\centering
	\includegraphics[width=\linewidth]{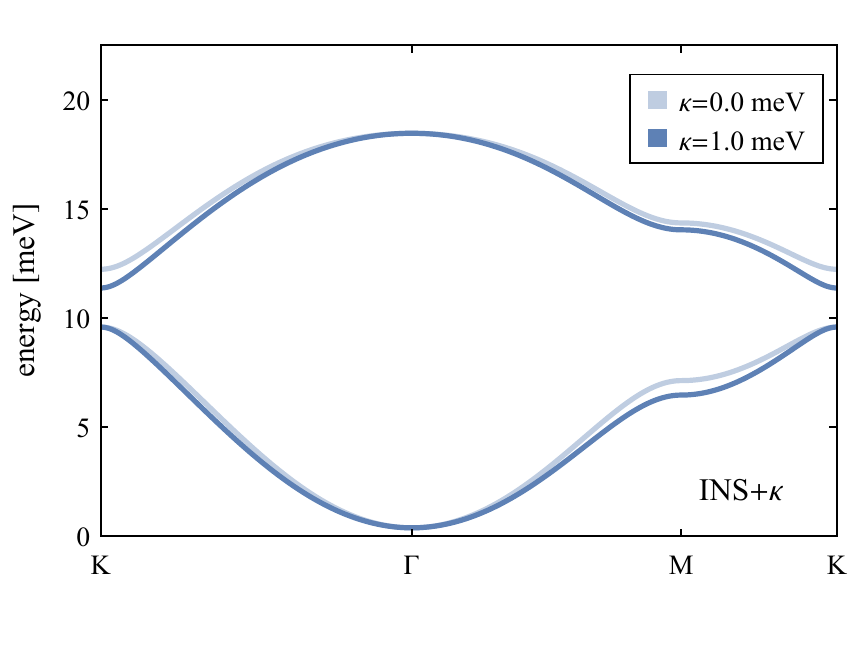}
	\caption{{\bf Kitaev-Heisenberg ambiguity} in the microscopic model for \ce{CrI3}. Plot shows the spin wave spectrum based on INS model parameters under deformed Heisenberg interaction $J_1'=J_1-\kappa$ and Kitaev interaction $K'=K+3\kappa$. The deformation of the spin wave spectrum under finite $\kappa$ is not linear. Up to $\kappa\approx 0.5~\mathrm{meV}$, no visible deformation is discernible. }
	\label{fig:sw:spinwave_INS_kappa}
\end{figure}

Yet, in addition to the visible differences in the spin wave spectrum, there exists a fundamental discrepancy which is more subtle: On the one hand, DFT computations predict a nearest neighbor antiferromagnetic Kitaev exchange constant of $K=0.6~\mathrm{meV}$ whereas the fit to experimental data which includes a DM term does not incorporate Kitaev exchange at all. 
On the other hand, it is possible to fit the experimental data with an entirely different microscopic model that has dominant ferromagnetic Kitaev interaction~\cite{Chen2020,Lee2020}; apparently, the model definition is ambiguous. 
Assuming ferromagnetic ground-state order, on a mean-field level, the nearest-neighbor Heisenberg coupling $J_1$ and the Kitaev exchange $K$ contribute an energy $E_\mathrm{mf}=S^2 (J_1+\frac{1}{3}K)$ per bond, where $S=3/2$ is the spin length~\cite{Lee2020}. 
We find that, as long as the mean-field energy scale $E_\mathrm{mf}$ is kept constant, it is possible to alter the relative weight of $J_1$ and $K$ without significantly impacting the spin wave spectrum. 
To illustrate this, we consider the microscopic model which has been fitted to inelastic neutron scattering data and which has exchange constants $(J_1,K)=(-2.11,0.0)~\mathrm{meV}$, among additional interactions detailed in Table~\ref{tab:dft:parameters}. 
We then deform the exchange constants $(J_1,K) \to (J_1',K') \equiv (J_1-\kappa,K+3\kappa)$, where $\kappa$ denotes the strength of the deformation, such that the mean-field energy scale $E_\mathrm{mf}$ is always preserved. 
Even a strong deformation of $\kappa=1~\mathrm{meV}$, which corresponds to $(J_1',K')=(-3.11,3.0)~\mathrm{meV}$, only has mild impact on the spin wave spectrum, as depicted in Fig.~\ref{fig:sw:spinwave_INS_kappa}. 

We therefore conclude that it is insufficient to simply fit a spin wave spectrum to neutron scattering data, since the fit cannot resolve the ambiguity between the Heisenberg and Kitaev exchange terms. 
Our DFT calculations, which are compatible with previous first principles calculations on more restricted model Hamiltonians~\cite{Xu2018}, suggest that the proposal of a Kitaev-dominated model~\cite{Lee2020} seems unlikely.
In particular, the relatively small nearest neighbor antiferromagnetic Kitaev interaction estimated from DFT has little impact on the spin-wave spectrum.
Nonetheless, it would be desirable to be able to probe the role of Kitaev interactions experimentally. 
In order to uncover properties of the model that are sensitive to the relative balance of Heisenberg and Kitaev exchange terms, we next discuss the impact of an external magnetic field on the spin wave spectrum.


\subsection{Spin waves of $\mathbf{CrI}_3$ in a magnetic field}

\begin{figure}
	\centering
	\includegraphics[width=\linewidth]{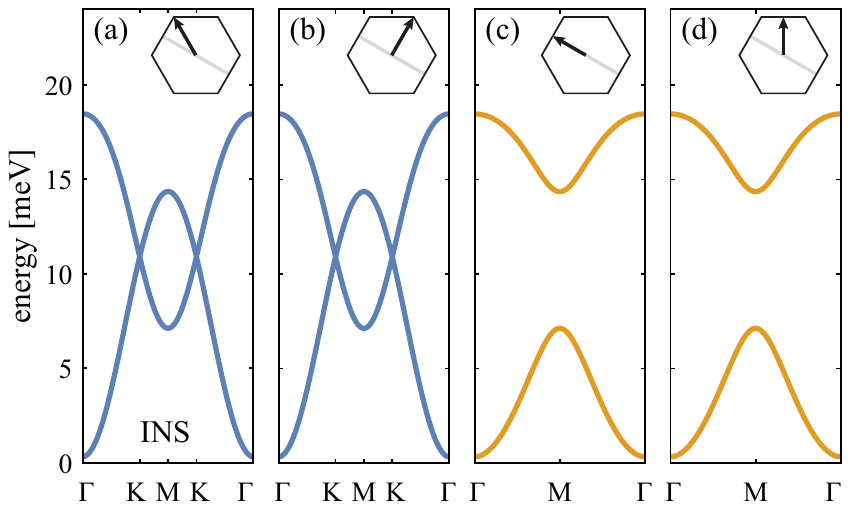}
	\caption{{\bf Dirac cones} in the spin wave spectrum of the INS model parameters in the high-field regime at $B=1.5B_c^\mathrm{2D}$. Spectrum is plotted along different momentum space cuts in panels (a)--(d). The cut directions are indicated in the insets by an arrow within the Brillouin zone; the spin polarization axis in reciprocal space is indicated by the gray line.}
	\label{fig:spinwave_INS_b_0.56_kappa_0.0_composite}
\end{figure}

\begin{figure}
	\centering
	\includegraphics[width=\linewidth]{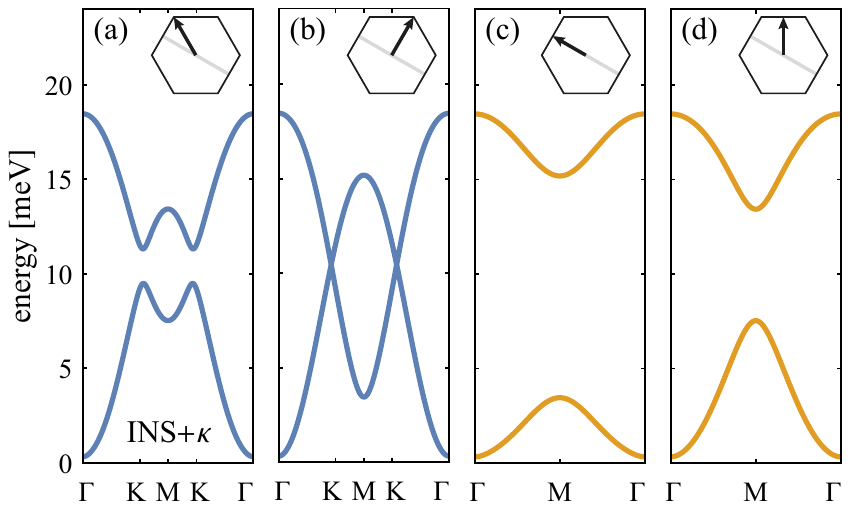}
	\caption{{\bf Dirac gap} in the spin wave spectrum of the INS model parameters with deformed Heisenberg interaction $J_1'=J_1-\kappa$ and Kitaev interaction $K'=K+3\kappa$ at $\kappa=1~\mathrm{meV}$ in the high-field regime at $B=1.5B_c^\mathrm{2D}$. Spectrum is plotted along different momentum space cuts in panels (a)--(d). The cut directions are indicated in the insets by an arrow within the Brillouin zone; the spin polarization axis in reciprocal space is indicated by the gray line. }
	\label{fig:spinwave_INS_b_0.56_kappa_1.0_composite}
\end{figure}

In this subsection, we discuss the changes which can be observed in the spin wave spectrum upon applying an external magnetic field to the model Hamiltonian Eq.~\eqref{eq:model:inplane}. 
The starting point for our discussion is the set of exchange constants which are fitted to the experimental neutron scattering data, listed in Table~\ref{tab:dft:parameters}. 
We then subject the model Hamiltonian to an in-plane magnetic field along the armchair direction $[\overline{1} 1 0]$. 
The field-coupling term $H_B = g \mu_B B \sum_i \hat{\mathbf{n}}_{\overline{1} 1 0} \mathbf{S}_i$ is added to the model Hamiltonian in analogy to our discussion of the magnetic phase diagram in Sec.~\ref{sec:mc}. 
By setting the field strength $B=4.8~\mathrm{T}$, which corresponds to approximately $1.5~B_c^\mathrm{2D}$ for this set of parameters, we ensure that the system is in its field-polarized phase. 

We reiterate that the model Hamiltonian used for these calculations (i.e., with exchange constants fitted to the inelastic neutron scattering data) does not contain any Kitaev interaction; the existence of a Dirac gap in the absence of an external magnetic field is solely due to the DM interactions.  
Now, with the magnetic moment polarized in-plane -- and thus perpendicular to the DM vector -- the DM interactions are effectively negated and the Dirac gap closes. 
The gapless Dirac cones in the spin wave spectrum are displayed in Figs.~\ref{fig:spinwave_INS_b_0.56_kappa_0.0_composite}a and~\ref{fig:spinwave_INS_b_0.56_kappa_0.0_composite}b for the two symmetry inequivalent directions defined by the external magnetic field. 
For completeness we also plot the spin wave spectrum along the two complementary high-symmetry directions, which cross the $M$-points of the BZ, see Figs.~\ref{fig:spinwave_INS_b_0.56_kappa_0.0_composite}c and~\ref{fig:spinwave_INS_b_0.56_kappa_0.0_composite}d. 
The key observation, which we point out here, is that the six-fold rotation symmetry remains intact and that the spin wave spectrum in the direction of all six Dirac points is equivalent.
Similarly, the spectrum along the directions of all six $M$-points is identical. 

\begin{figure}
	\centering
	\includegraphics[width=\linewidth]{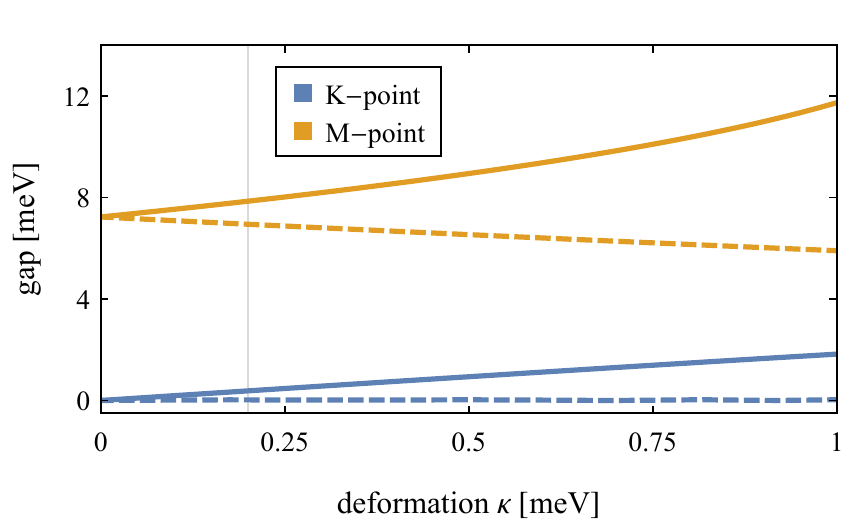}
	\caption{{\bf Gap sizes} at the $K$-points and the $M$-points for the INS model parameters with deformed Heisenberg interaction $J_1'=J_1-\kappa$ and Kitaev interaction $K'=K+3\kappa$ in the high-field regime at $B=1.5B_c^\mathrm{2D}$. The gaps at the $K$-points (blue lines) and $M$-points (yellow lines) become anisotropic at finite $\kappa$. Solid lines correspond to the momentum direction indicated in panels~(a) and~(c) of Fig.~\ref{fig:spinwave_INS_b_0.56_kappa_1.0_composite}, dashed lines correspond to the momentum direction in panels~(b) and~(d), respectively. The gray line indicates Kitaev interaction $K'=0.6~\mathrm{meV}$, the value we find in our DFT calculations.}
	\label{fig:spinwave_INS_b_0.56_kappa_dependence}
\end{figure}

Let us now explore the changes that manifest when the opening of a Dirac gap is no longer exclusively due to the DM interaction. 
For this purpose, we re-introduce the Heisenberg-Kitaev deformation $(J_1,K) \to (J_1',K') \equiv (J_1-\kappa,K+3\kappa)$, which was discussed in the previous subsection. 
Kitaev interactions (in the absence of an external magnetic field) give rise to a finite Dirac gap, but the gap remains small unless the Kitaev interaction becomes the dominant term in the model~\cite{Lee2020,Chen2020}. 
With the Heisenberg-Kitaev deformation in place, for $\kappa\neq 0$, we no longer necessarily expect that the Dirac gap closes when the system is in its field-polarized phase. 
Indeed, we demonstrate in Fig.~\ref{fig:spinwave_INS_b_0.56_kappa_1.0_composite}a an example where for $\kappa=1~\mathrm{meV}$ the gap remains open. 
However, this does not mean that the role of Kitaev interactions in unaffected by the magnetic field. 
In fact, its role is highly dependent on the field: As pointed out in Ref.~\cite{Chen2021}, in the absence of DM interactions the six-fold rotational symmetry of the Dirac points is broken. 
The symmetry breaking persists also in the presence of finite DM interactions, as demonstrated in Fig.~\ref{fig:spinwave_INS_b_0.56_kappa_1.0_composite}b; the gap remains finite only at four of the Dirac points, while the remaining two become gapless. 

In analogy to the lifting of degeneracies among the Dirac points, the spin wave spectrum around the $M$-points of the BZ also splits into two symmetry inequivalent classes. 
The band gap at the two $M$-points that lie in the direction of the magnetic field slightly increases, whereas the band gap at the four $M$-points with finite perpendicular components to the magnetic field direction decreases, as illustrated in Figs.~\ref{fig:spinwave_INS_b_0.56_kappa_1.0_composite}c and~\ref{fig:spinwave_INS_b_0.56_kappa_1.0_composite}d. 
This effect has implications for our understanding of the underlying microscopic model. 
Previously, in the absence of a magnetic field, we established that a reweighting between the nearest neighbor Heisenberg and Kitaev exchange constants has negligible impact on the spin wave spectrum as long as the mean-field energy scale $E_\mathrm{mf}=S^2 (J_1+\frac{1}{3}K)$ remains constant. 
Now, we have identified an observable that can probe the existence of Kitaev interactions in the material. 
We now make the probe more quantitative. 
To this end, we calculate the splitting of gaps between the two classes of symmetry inequivalent $K$-points ($M$-points) as a function of the deformation parameter $\kappa$. 
The splitting scales approximately linearly with $\kappa$, as demonstrated in Fig.~\ref{fig:spinwave_INS_b_0.56_kappa_dependence}. 
We shall mention that the splitting is inherently driven by the exchange constants based under the assumption of a fully polarized spin configuration; further increasing the magnetic field strength does not increase the gap splitting. 
At $\kappa=0.2~\mathrm{meV}$, which is the amount of deformation needed for the model with exchange constants extracted from neutron scattering to incorporate the amount of Kitaev interaction that we predict in our DFT calculations (cf. Table~\ref{tab:dft:parameters}), we find that the splitting between $K$-points is $\Delta_K=0.36~\mathrm{meV}$ and the splitting between $M$-points is $\Delta_M=0.91~\mathrm{meV}$. 
Such an anisotropy in the spin wave spectrum could be probed in high-field neutron scattering experiments.


\section{Discussion}
\label{sec:discussion}

In this work, we performed first-principle calculations to predict a complete set of exchange constants for a microscopic model of \cri{}. 
Our calculations simultaneously include the effects of Kitaev interaction and DM interaction and therefore mark a significant extension of earlier work: 
Previous calculations only separately addressed the role of Kitaev and DM interactions, yet both are suited to model a gap in the magnon band structure that has been observed experimentally at the Brillouin zone corners ($K$-points) and that is crucial to capturing the spin dynamics of \cri{}~\cite{Chen2020,Chen2021}.
Including both types of interactions in our model, we were able to show that the Dirac gap is likely driven by DM interactions, with only minor contribution from Kitaev interaction. 
A previously proposed microscopic model in which the gap is driven by dominant Kitaev exchange interactions~\cite{Lee2020} can be ruled out based on our calculations. 
We have found that the magnetic transition temperature is greatly overestimated by the earlier mean-field approach~\cite{Lee2020} which appears to have led to an overemphasis of the Kitaev interaction. 

Furthermore, we performed classical Monte Carlo simulations to determine the phase diagram of our model Hamiltonian as a function of temperature and an external in-plane magnetic field. 
At zero field, we demonstrated that the model yields an ordering temperature of $T_c^\mathrm{2D}=40.6~\mathrm{K}$ in the monolayer limit, which is in good agreement with the experimentally observed value $T_{c,\mathrm{expt}}^\mathrm{2D}=45~\mathrm{K}$. 
As such, our model simultaneously captures the excitation spectrum of \cri{} as well as its thermal properties. 

The exchange constants that comprise our model are similar to a set of interaction parameters that has been extracted from inelastic neutron scattering data~\cite{Chen2021}, yet we point out that the role of sub-dominant Kitaev interactions has not been tested in this earlier work. 
We showed that performing such tests is challenging, because it is possible to systematically modify the Kitaev interactions of the microscopic model in a way that has little impact on the magnon band structure -- and thus on the neutron scattering experiments --, leaving an ambiguity in their interpretation.  
However, we demonstrated that the ambiguity can be lifted by probing the band structure in a finite in-plane magnetic field. 
In such setting, when the spins are fully polarized in the honeycomb plane, the typically dominant contribution of DM interactions towards the opening of a Dirac gap is suppressed and the effect of Kitaev interactions becomes measurable. 
We showed that the degeneracy of the six Dirac gaps is lifted in the presence of Kitaev interactions, thus breaking the six-fold rotation symmetry, and that the degree of splitting systematically depends on the strength of the Kitaev interaction. 
It would therefore be desirable to perform higher resolution INS experiments to probe the anisotropy of the spin wave spectrum of \cri{} in the presence of a magnetic field.


\vspace{-11.1pt}
\begin{acknowledgments}
We thank Pengcheng Dai and Lebing Chen for useful discussions.
The numerical simulations were performed on the Cedar cluster, hosted by WestGrid and Compute Canada. 
Monte Carlo simulations made use of the SpinMC.jl Julia package~\cite{SpinMC}, and spin wave calculations were performed with the SpinW software~\cite{Toth2015}. 
FLB and AP acknowledge funding from NSERC of Canada. 
 I.D. thanks Department of Science and Technology, Technical Research Centre (DST-TRC)
and Science and Engineering Research Board (SERB) India
(Project No. EMR/2016/005925) for support.
 FU and NT thank the Center of
Emergent Materials, an NSF MRSEC under award number DMR-2011876 for support.
\end{acknowledgments}


\bibliography{cri3}

\end{document}